# Atmospheric pressure intercalation of oxygen via wrinkles between graphene and a metal

*Amina Kimouche,*[1] *Olivier Renault,*[2] *Sayanti Samaddar,*[1] *Clemens Winkelmann,*[1] *Hervé Courtois,*[1] *Olivier Fruchart,*[1] and *Johann Coraux*[1,*]

[1]Institut Néel, CNRS et Université Joseph Fourier, BP 166, Grenoble Cedex 9, 38042, France
[2]CEA, LETI, MINATEC Campus, 17 rue des Martyrs, 38054 Grenoble Cedex 09, France
* johann.coraux@grenoble.cnrs.fr

Changes of paradigms, in terms of new functionalities, architectures, and performances, are foreseen with graphene, an atomically thin sheet of carbon atoms in a honeycomb lattice. These prospects are urging the development of efficient production methods.[1] Preparation by chemical vapor deposition (CVD), in this respect, has reached such maturity that graphene now appears as an alternative to indium tin oxide as a transparent conductive electrode[2] or to Si and II-IV semiconductors in high-frequency electronics.[3] Intercalation of species between the metallic substrate needed for CVD and graphene, a method known since the 1980's,[4] is an efficient and versatile way to achieve quasi free-standing graphene[5] and to engineer the properties of graphene, for instance to induce electronic band-gaps,[6] magnetic moments,[7] and strains.[8] Dual intercalation, of Si and O, even showed great promise for the transfer-free preparation of graphene-on-oxide field effect transistors.[9]

Despite the numerous reports devoted to graphene/substrate intercalated systems, two key questions remain open. First, the surmised role of defects as pathways for intercalation has only been established, yet partially in some cases, for a few defects, namely graphene free edges[5] and point defects.[10,11] Unveiling other intercalation pathways will help better envisioning the full potentialities of intercalation for building up advanced graphene-based hybrids. Second, all studies of intercalation reported thus far were performed under ultra-high vacuum (UHV). While this approach offers optimum control over the processes, it is a prohibitively costly one in the view of the production of graphene decoupled from its substrate. While atmospheric pressure intercalation would be desirable, intercalation may proceed differently due to the markedly different conditions.

In this article we address these two questions, by studying high quality graphene prepared by CVD on Ir(111) thin films. Iridium is one of those transition metals, like Cu, Pt, and Au, that weakly interact with graphene.[12] Unlike on the latter three metals however, most defects and notably grain boundaries can be avoided on Ir(111),[13,14] so that their role in intercalation[10] can be ruled out. Correlating complementary microscopic analysis, we show that exposure to air leads to the formation of an ultrathin intercalated oxide modifying the weak graphene-Ir interaction and generating variations of the charge density in graphene. We find that a common kind of defect, local delaminations of graphene from its substrate, so-called wrinkles, forming during cool down to room temperature after CVD due to the mismatch of thermal expansion coefficient of graphene and its substrate, are pathways for a slow intercalation of oxygen species.

Under UHV, prior to exposure to air the graphene-covered regions of the sample exhibit a typical flat topography, except for wrinkles and a periodic corrugation associated with the graphene/Ir(111) moiré[12] (Supplementary Information (SI), Figure S1). The situation is markedly different after exposure to air, as shown by atomic force microscopy (AFM). Inspecting phase contrast AFM images (SI, Figure S2) allows one to identify graphene-free and graphene-covered regions.[15] Figure 1a reveals that (i) graphene-free regions (named A hereafter) protrude out of the surface, (ii) around graphene edges and along wrinkles, flat ribbons (named B hereafter) whose height is substantially larger (see below) than the Ir atomic step edges, are found. The moiré observed with scanning tunneling microscopy (STM) under UHV before air exposure is no more visible with STM.[16]

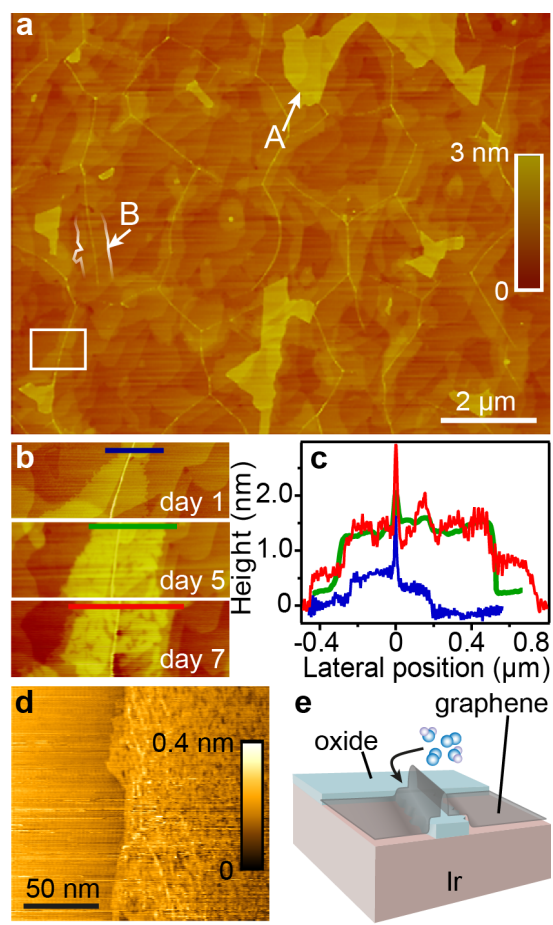

*Figure 1.* Oxidation of iridium on graphene-free and graphene-covered regions. a) AFM image of graphene/Ir(111) after exposure to air, exhibiting Ir oxide on graphene-free regions (labeled "A") and intercalated between graphene and Ir, in the form of ribbons along the wrinkles in graphene (labeled "B"). b) Close-up of the region framed in a), as a function of time. c) Height profiles along the lines marked in b). "0" marks the position of the wrinkle. d) STM topograph of a graphene-covered region comprising a type-B region (right part). e) Cartoon summarizing the effect of atmospheric molecules (sketched with balls).

While the height (typically 3 nm in average, relative to graphene) of type A-regions does not evolve with time, the height of type B-regions, as well as their lateral expansion, does (Figure 1b,c). After a few days the height and width of type B-regions do not increase much and reaches about 1.5 nm and a few 100 nm respectively. The surface of graphene appears rougher on type-B regions than on other graphene regions (Figure 1d). What is the nature of the height variations upon exposure to air? Volume expansion of surfaces due to their oxidation is a well-known effect in AFM.[17] The presence of type A-regions is interpreted in this light: in the absence of a protecting layer hindering surface air oxidation, an Ir oxide rapidly forms.[18,19] This oxide has a low conductivity, compared to the rest of the sample (SI, Figure S3). Type B-regions, which are covered with graphene, are tentatively identified at this point as Ir oxide ultrathin films progressively forming upon the reactive intercalation of oxygen.

In order to confirm this scenario, we performed spatially-resolved measurements of the work-function (WF).[20] PEEM spectroscopic images at the photoemission threshold exhibit a contrast inversion as a function of the secondary electron energy, indicative of a strong WF contrast across the surface (Figure 2a and S5). WF maps reveal two kinds of regions (Figure 2b,c): one with a 4-4.1 eV WF, the other with a mean 4.5 eV WF and variations of approximately 0.1 eV about this values across micrometer distances. The first value is close to the 4.2 eV one reported for $IrO_2$,[18] and thus presumably corresponds to type A-regions in Figure 1. The second value is similar to that reported for graphene/Ir(111) (Ref. [21]) and thus signals graphene-covered regions, the spatial fluctuations in their WF seemingly corresponding to surface inhomogeneities, with length-scale compatible with the distance between intercalated (type B) ribbons.

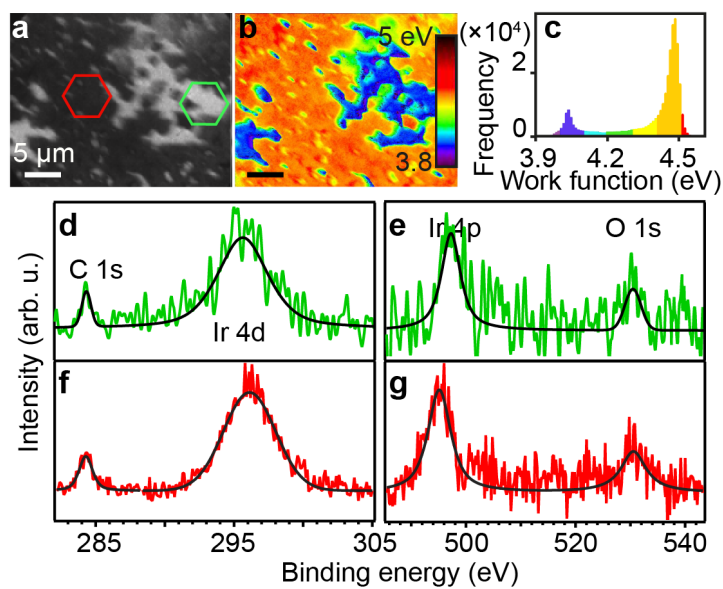

*Figure 2.* Surface electronics and elemental composition. a) UV-excited PEEM spectroscopic image (4.4 eV secondary electrons) of a surface area comprising graphene-free (type A-regions) and graphene-covered regions, including type B-regions (oxide ribbon intercalated below graphene). b) Corresponding WF mapping of the region in a). c) WF distribution of b). d-g) Background-subtracted XPS micro-spectra of Ir, C, and O core levels over two areas of interest marked in a) (colored spectra are data obtained from the hexagonal frames in a) having the corresponding color; black curves are fits to the data).

The picture sketched above is further supported by chemical analysis of the regions with different WF, thanks to area-selected X-ray photoemission performed with the same PEEM set-up operated in the micro-spectroscopic mode.[20] We find that the intensity ratio between C 1s and Ir 4d peaks (Figure 2d,f) is 12 % in the high WF regions, and about twice as low on regions with prominently low WF. We note that due to the size of the iris aperture for area selection (sketched in Figure 2a), low WF regions also comprise a contribution (~ 30% here) of high WF regions. We deduce that low WF regions are graphene-free. An O 1s peak is detected in all regions (Figure 2e,g). Though the statistics of our measurement does not allow to map out spatial inhomogeneities of this species, combining the information, gained from the WF and chemical analysis, we can conclude that low WF regions are graphene-free, type A-regions ($IrO_2$), and that high WF regions are graphene-covered and comprise oxygen species. The latter are inhomogeneously distributed over the surface, type B-regions (oxide ribbons intercalated along wrinkles), and possibly also, between these regions, a disordered layer of small oxygen-containing molecules (*e.g.* water, dioxygen) intercalated below graphene.

Final version of the manuscript at http://dx.doi.org/10.1016/j.carbon.2013.10.033

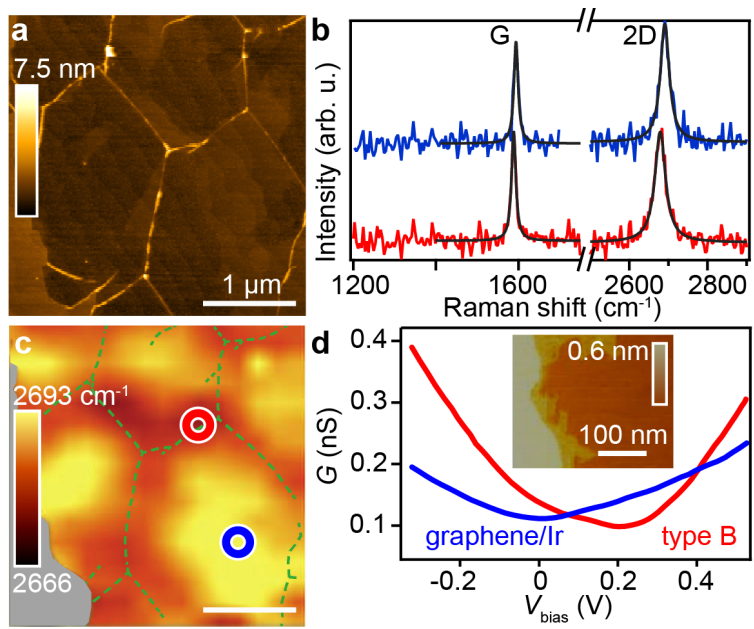

*Figure 3.* Change of the properties of graphene upon oxide intercalation. a) AFM image of graphene/Ir. b) Raman spectra acquired on the two spots marked in c). c) Map of the position of graphene's 2D band Raman shift, in the same area as a). The grey region is graphene-free, doted lines are guides for the eyes showing the position of the wrinkles apparent in a). d) Local conductance (G) as a function of tip-sample bias ($V_{bias}$) averaged on the left region (red, type-B) region, and on the right (blue, graphene on Ir(111) without intercalated ribbon) of the STM image shown in inset.

At this point we summarize (Figure 1e) the effect of exposure to atmospheric environment at room temperature. Without holes in the graphene sheet, no oxidation of the Ir surface occurs. Graphene-free regions are rapidly oxidized, on a time scale below 1 h. Graphene-covered regions are also oxidized, but on a much larger timescale, upon the slow intercalation of oxygen species at graphene edges entering wrinkles at their free end, which act as a tunnel for the transport of species across at least few micrometer distances. We note that the intercalation does not occur in case of high quality graphene fully covering the Ir(111) surface (as also noticed in Ref. [22]), *i.e.* when wrinkles do not have an open end.

In the remaining we focus on the type B-regions, more precisely on the influence of the intercalated oxide on the properties of graphene. Spatially-resolved Raman scattering measurements were performed at the exact location where AFM (Figure 3a) was also performed. No defect-allowed D mode is observed (Figure 3b and S4), which confirms the uniform high quality of the graphene-covered regions. In the vicinity of the wrinkles, graphene's G and 2D modes are found at 1588±3 and 2679±3 cm$^{-1}$ respectively; they are located at 1594±3 and 2691±3 cm$^{-1}$ far from the wrinkles (Figure 3b). Though the G and 2D mode softening, about 6 and 10 cm$^{-1}$, correlate with the position of the wrinkles (Figure 3c and S4), they are actually not a direct manifestation of graphene's deformation or decoupling from its support in the wrinkles. The G and 2D modes indeed have single-lorentzian profiles, in other words, they do not comprise two spectral components, which would each arise from "flat" and wrinkled graphene. A significant contribution from the wrinkles is actually not expected, given that the number of the atoms which they encompass is typically 100 times smaller than that illuminated by the laser spot (300 nm diameter). The observed blue-shifts are thus ascribed to the presence of the intercalated oxide ribbon. Strain and changes in electronic density can both be invoked in principle to explain G and 2D mode shifts. Rough estimates based on reports addressing specifically each effect yield order of magnitudes, 0.1 % and 10$^{12}$ cm$^{-2}$ for strains[23] and electronic density change[24] respectively. The former could result from the stretching of graphene upon intercalation;[8] the latter could be associated with variations of the charge density in graphene set by the variation of the distance between graphene and the metal substrate due to the presence of the intercalated oxide.[16]

Scanning tunneling spectroscopy was employed to test which effect (strain or charge transfer) actually prevails. The local density of state (LDOS, which is proportional to the local conductance measured with the STM tip) is markedly different on graphene regions with and without an intercalated oxide ribbon (Figure 3d). Concluding from the shape of the LDOS whether the graphene/support interaction is stronger or weaker is in our opinion daring, given the different shapes reported by different groups on identical graphene systems.[25,26] We shall only mention that the graphene/Ir interaction, which is known to be weak[12] can reasonably be expected to be even weaker in the presence of an interaction oxide, as is the case in graphene/Ru(0001).[9] Assuming that the LDOS minimum tracks the neutrality point in graphene,[27,28] we first observe that the Dirac point for graphene on Ir(111) is located closer to zero than in *in situ* UHV experiment.[12] The difference is ascribed to a transfer of electrons from below and above graphene. The latter may include the effect of adsorbates on graphene. The former includes the effect of a disordered layer of molecules possibly intercalated between graphene and Ir, the presence of which would explain both the occurrence of a strong Raman signal absent in as-grown graphene/Ir(111),[29] and the absence of the moiré pattern. Changes in the position of the Dirac point, of a few 100 meV, are observed between graphene with and without an oxide below it. This translates into variations of charge densities of the order of $10^{12}$ cm$^{-2}$, which account for the observed Raman shift changes. We deduce that the latter originate prominently from charge density effects induced by intercalation.

In conclusion, we discovered that wrinkles play a central role in the diffusion of oxygen species in between graphene and its Ir(111) substrate under atmospheric conditions, by letting species penetrating below graphene from their open end. These diffusing species react with Ir and slowly form oxide ribbons whose thickness reaches 1.5 nm, expanding across a self-limited width of a few 100 nm. The effect was also observed recently in our laboratory on another related system, graphene on Cu foils, and is thus of general importance in CVD-produced graphene. We found strong evidence that the intercalation of the oxide ribbons yields modulation of few $10^{12}$ cm$^{-2}$ of the charge density in graphene and strongly modifies the graphene-support interaction. Our findings highlight a cheap and easy way of intercalating thin oxides between graphene and its substrate, and on the contrary, the crucial need for eliminating holes in graphene on metals with which it weakly interacts, in the view of any application taking benefit of graphene as a barrier coating to oxidation. More advanced structures, with in-plane variations of the nature of the intercalants across 100 nm distances, could be designed by further processing. This opens the way to the bottom-up fabrication of high quality charge-modulated graphene, for instance in the view of smart devices exploiting the principles of optics for guiding charge carriers.[30,31]

**Experimental section**

*Graphene preparation*: Graphene was prepared by CVD at 1300 K with ethylene, in a UHV system with $10^{-10}$ mbar base pressure, on Ir(111) thin films previously deposited on C-plane sapphire wafers.[34] The growth was stopped after few minutes of exposure to a $10^{-8}$ mbar partial pressure of ethylene in order to achieve a partial coverage of graphene.

*Microscopic characterizations:* AFM was performed in tapping mode in atmospheric pressure. Raman scattering measurement were done in atmospheric pressure with a WITec spectrometer coupled to an excitation laser with 532 nm wavelength and 2 mW/cm² power, and a 100× objective lens allowing a spatial resolution of about 300 nm. PEEM measurements were done under UHV after mild degassing (500 K) of the sample for 1 h, using a NanoESCA spectromicroscope (Omicron NanoTechnology). For the WF analysis, a UV Hg light source (4.9 eV) was used; the lateral resolution was better than 150 nm and the typical sensitivity of the WF determination was 20 meV. For the photoemission micro-spectra, a focused X-ray source (1486 eV) was illuminating the typical 35 μm field of view, and area selection over 5 μm-wide regions was performed thanks to a field aperture located in the first intermediate image plane of the PEEM optics; the overall energy resolution was 0.8 eV.

*STM/STS measurements***:** STM measurements were performed at room temperature. The spectra displayed are the average over 132 spectra, each obtained over 132 different points separated by 3.1 nm in each of the regions. STM images were acquired with 100 pA and 800 mV tunneling current and bias respectively.

**Acknowledgements**
We thank Valérie Guisset and Philippe David for valuable support during the setting-up of the UHV experiments, and Nedjma Bendiab for fruitful discussions. Research supported by EU contract NMP3-SL-2010-246073 "GRENADA", and the French ANR contract ANR-2010-BLAN-1019-NMGEM.